# Advancing quantitative understanding of self-potential signatures in the critical zone through long-term monitoring

Kaiyan Hu[1,2], Damien Jougnot[3], Qinghua Huang[1], Majken C. Looms[4], Niklas Linde[2,*]

[1]Department of Geophysics, Peking University, 100871 Beijing, China

[2]Institute of Earth Sciences, University of Lausanne, 1015 Lausanne Switzerland

[3]Sorbonne Université, CNRS, EPHE, UMR 7619 Metis, F-75005, Paris, France

[4]Department of Geosciences and Natural Resource Management, University of Copenhagen, Øster Voldgade 10, 1350 Copenhagen, Denmark

[*] Corresponding author. E-mail address: niklas.linde@unil.ch

**Author contact:**

Kaiyan Hu: huk@pku.edu.cn; kaiyan.hu@unil.ch

Damien Jougnot: damien.jougnot@upmc.fr

Qinghua Huang: huangq@pku.edu.cn

Majken C. Looms: mcl@ign.ku.dk

Niklas Linde: niklas.linde@unil.ch

**ABSTRACT**

The self-potential (SP) method is a passive geophysical technique, which may offer insights about water and ionic fluxes in the vadose zone. The main obstacles presently prohibiting its routine use in quantitative vadose zone hydrology are the superposition of signals arising from various source mechanisms, difficult-to-predict electrode polarization effects that depend on electrode design and age, as well as water saturation, pore water chemistry, clay content, and temperature in the immediate vicinity of the electrodes. We present a unique long-term SP monitoring experiment focusing on the first four years of data acquired at different depths in the vadose zone within the HOBE hydrological observatory in Denmark. Using state-of-the-art SP theory combined with flow and transport simulations, we attempt to replicate the observed data and suggest reasons for observed discrepancies. The predictions are overall satisfactory during the first six months of monitoring after which both the patterns and magnitudes of the observed data change drastically. Our main observations are (1) that predicted SP magnitudes are strongly sensitive to how the effective excess charge (or alternatively, the voltage coupling coefficient) scales with water saturation implying that continued research is needed to build more accurate models of electrokinetic phenomena in unsaturated conditions, (2) that significant changes in electrode polarization occur in the shallowest electrodes at time scales of a year, most likely due to desaturation by capillarity of the fluids filling the electrodes, suggesting that electrode effects cannot be ignored and that explicit electrode modeling should be considered in future monitoring studies, and (3) that multi-rate mass transfer and reactive transport modeling, with specific emphasis on fluid-mineral interactions, are needed to better predict salinity and pore water conductivity. We hope to stimulate other researchers to test new SP modeling approaches and interpretation strategies against these data by making the SP and complimentary (temperature, dielectric constant, potential/actual evapotranspiration, precipitation) data time-series available.

**Keywords:** Hydrogeophysics; Long-term monitoring; Self-potential; Vadose zone; Electrode effects; Solute transport.

# 1. Introduction

Self-potential (SP) signals are mainly generated by naturally-occurring electrokinetic, electrochemical and bioelectrical source mechanisms (Revil and Jardani, 2013). In the SP method, passive measurements with non-polarizing electrodes are used to map or monitor voltages on the ground surface or in boreholes. Due to a rich set of natural source mechanisms, the SP method has been used to study a wide range of hydrological and critical zone processes. Most studies have based their interpretations on the electrokinetic effect, which originates when water flowing in porous media drags excess electric charges in the electrical double layer to produce the so-called streaming potential. SP signals attributed to streaming potentials have been used to characterize hydraulic parameters (e.g., Zlotnicki and Nishida, 2003; Jardani et al., 2009; Jougnot et al., 2012; Soueid Ahmed et al., 2014), landslides (Lapenna et al., 2003) and seepage from dams (e.g., Ikard et al., 2014; Rittgers et al., 2014; Soueid Ahmed et al., 2019). SP signals of electrochemical origin arise in the presence of ionic concentration gradients in the pore water or when redox reactions are facilitated by electronic conductors. Self-potential signals of electrochemical origin have been used to constrain saline tracer and contaminant transport (e.g., Maineult et al., 2005; Martínez-Pagán et al., 2010; Jougnot et al., 2015), sea water intrusion (e.g., Graham et al., 2018; MacAllister et al., 2018), to locate the corrosion of buried metallic bodies (e.g., Castermant et al., 2008; Rittgers et al., 2013) and to invert for redox maps associated with contaminant plumes (Linde and Revil, 2007). In the SP literature, it is often tacitly ignored that recorded voltages do not only comprise information about subsurface processes of interest such as water and solute fluxes in the vadose zone. The recorded voltages are also adversely affected by electrode-related effects (Petiau and Dupis, 1980; Perrier et al., 1997; Petiau, 2000; Jougnot and Linde, 2013) and external sources originating from telluric currents and man-made electrical and electromagnetic installations (Perrier et al., 1997; Park et al., 2007; DesRoches and Butler, 2016).

Self-potential monitoring data contain transient effects, as well as more periodic short- and long-term variations of different origins. In the literature, SP data are often interpreted in terms of one process of primary interest, while other possible effects and time-scales are ignored or filtered out. Despite a large number of published SP studies, only few have concerned longer-term monitoring (Perrier et al., 1997; Doussan et al., 2002; Voytek et al., 2019). Here, we investigate four years of SP monitoring data acquired with permanently installed electrodes at different depths within the vadose zone at the HOBE hydrological observatory (www.hobecenter.dk) in Denmark. Our emphasis is on understanding the main signatures of the observed SP data and to highlight effects that currently limit our ability to make quantitative predictions about vertical water and solute fluxes in the vadose zone. Instead of filtering out unwanted effects such as electrode offsets and quasi-linear drifts, we attempt to use the SP record in its entirety to truly understand to what extent the SP method can be used as a predictive tool in vadose zone and critical zone science. For instance, can SP monitoring data be used in a reliable manner to estimate *in situ* water fluxes in the vadose zone? Or more simply, can SP data be used to determine if water fluxes at a given time and depth are directed upward or downward? We contrast the measured SP data with state-of-the-art SP numerical modeling that accounts for electrokinetic and electrochemical contributions, as well as electrode effects. In an effort to engage the SP community in research towards an improved mechanistic and quantitative understanding of SP signals in the critical zone, we make the full SP and complimentary (dielectric constant, temperature, potential and actual evapotranspiration, precipitation) data record available.

## 2. Methodology

Section 2.1 introduces basic self-potential theory followed by more detailed sections describing the electrokinetic contribution (section 2.1.1), the electro-diffusive contribution (section 2.1.2), and electrode-effects (section 2.1.3) leading to the total SP response (section 2.1.4). Section 2.2 introduces the hydrodynamic characterization and modeling that is needed input for self-potential modeling in the vadose zone. Water flow and solute transport (section 2.2.1) is first introduced before explaining how *in situ* field data are used to infer vadose zone properties by inversion (section 2.2.2).

*2.1. Self-potential theory*

Below we describe the theory used to perform numerical simulations of SP signals in the vadose zone. In this treatment, we exclude the redox contribution since we do not expect electronic conductors at the HOBE site.

In the absence of external current sources, we have:

$$\nabla \cdot \mathbf{J_{tot}} = \nabla \cdot (\mathbf{J_c} + \mathbf{J_{ek}} + \mathbf{J_{ed}} + \mathbf{J_{ee}}) = 0, \tag{1}$$

where the total electric current density $\mathbf{J_{tot}}$ is the sum of the conductive current density $\mathbf{J_c}$ (A/m$^2$), the electrokinetic current density $\mathbf{J_{ek}}$ (A/m$^2$) and the electro-diffusive current density $\mathbf{J_{ed}}$ (A/m$^2$) and electrode-related contributions $\mathbf{J_{ee}}$ (A/m$^2$). The conductive current density $\mathbf{J_c}$ (A/m$^2$) obeys Ohm's law:

$$\sigma(\theta_w)\mathbf{E} = \mathbf{J_c}, \tag{2}$$

where $\mathbf{E}$ (V/m) is the electric field, which is the negative gradient of electrical potential $\varphi$ (V), $\sigma(\theta_w)$ (S/m) is the effective electrical conductivity of the soil. For coarser soils without fines, we have (Archie, 1942):

$$\sigma(\theta_w) = F^{-1}\sigma_w S_w^{\,n}, \tag{3}$$

where $S_w = \frac{\theta_w}{\phi}$ is the water saturation, $\theta_w$ is the water content, $F = \phi^{-m}$ is the formation factor, $\phi$ is the porosity, $m$ is the cementation exponent and $n$ is the saturation exponent. For soils containing fines, the relationship between bulk electrical conductivity and pore water electrical conductivity is rather complex and non-linear (e.g., Revil et al., 2019). Assuming that the pore water is a NaCl-solution, the electrical conductivity of pore water $\sigma_w$ (S/m), the temperature of water $T$ (°C) and solute concentration $C_w$ (mol/L) are related by (Sen and Goode, 1992):

$$\sigma_{\text{w}} = (5.6 + 0.27T - 1.5 \times 10^2 T^2)C_{\text{w}} - \frac{2.36 + 0.099T}{1 + 0.214 C_{\text{w}}} C_{\text{w}}^{\frac{3}{2}}. \tag{4}$$

*2.1.1. Electrokinetic contribution*

The electrokinetic current density depends on the water flux **u** and the effective excess charge density $\widehat{Q_{\text{v}}^{\text{eff}}}(\theta_{\text{w}}, C_{\text{w}})$ (C/m³) of porous media:

$$\mathbf{J}_{\text{ek}} = \widehat{Q_{\text{v}}^{\text{eff}}}(\theta_{\text{w}}, C_{\text{w}})\mathbf{u}, \tag{5}$$

where $\widehat{Q_{\text{v}}^{\text{eff}}}(\theta_{\text{w}}, C_{\text{w}})$ varies, for a given soil, with the water content $\theta_w$ and solute concentration $C_{\text{w}}$. It might also exhibit a hysteretic behavior that we ignore in the present treatment (Revil et al., 2008). The main challenge in deriving the effective excess charge is which it is a flux-averaged property (Linde, 2009). In this work, we will mainly rely on the water retention (WR) formalism introduced by Jougnot et al. (2012) in which the soil is conceptualized as a bundle of capillaries that can be either filled with water or air (Jackson, 2008). In this approach, $\widehat{Q_{\text{WR}}^{\text{eff}}}(\theta_{\text{w}}, C_{\text{w}})$ is derived from the water retention function by integrating the effective excess charge density $\widehat{Q_{\text{v}}^{\text{eff},r}}(r, C_{\text{w}})$ of each saturated capillary with radius $r$ over all capillaries in the soil:

$$\widehat{Q_{\text{WR}}^{\text{eff}}}(\theta_{\text{w}}, C_{\text{w}}) = \frac{\int_{r_{\min}}^{r_{\max}} \widehat{Q_{\text{v}}^{\text{eff},r}}(r, C_{\text{w}}) v^r(r) f_{\text{WR}}(r) dr}{\int_{r_{\min}}^{r_{\max}} v^r(r) f_{\text{WR}}(r) dr}, \tag{6}$$

where $v^r(r)$ is the pore water flux in a capillary and $f_{\text{WR}}(r)$ is the capillary size distribution derived from the relationship between pore pressure and saturation under the van Genuchten water retention model (van Genuchten, 1980). An alternative for calculating the capillary size distribution is the relative permeability (RP) method $f_{\text{RP}}(r)$, which derives the saturation from the relative permeability function (Jougnot et al., 2012). For the WR-approach, the relative excess charge density $\widehat{Q_{\text{WR}}^{\text{eff,rel}}}(\theta_w, C_w)$ curves defined by the ratio of effective excess charge density with saturated excess charge density $\frac{\widehat{Q_{\text{WR}}^{\text{eff}}}(\theta_w, C_w)}{\widehat{Q_{\text{WR}}^{\text{eff,sat}}}(C_w)}$ depend on the shape of the water retention curves ($\alpha_{\text{VG}}$ and $n_{\text{VG}}$) of different materials. This approach has been shown to predict the evolution of the effective excess charge density with respect to the water content (e.g., Jougnot and Linde, 2013; Voytek et al., 2019). Here, $\alpha_{\text{VG}}$ (m$^{-1}$) refers to the inverse of the air entry pressure and $n_{\text{VG}}$ (-) is related to the pore-size distribution. Instead of using Eq. (6) and in perfect analogy with the treatment of hydraulic conductivity in unsaturated media, we scale this relative excess charge density with laboratory or field data at saturated conditions. This leads to

$$\widehat{Q^{\text{eff}}}(\theta_w, C_w) = \widehat{Q_{\text{WR}}^{\text{eff,rel}}}(\theta_w, C_w)\, \widehat{Q^{\text{eff,sat}}}(C_w), \qquad (7)$$

where $\widehat{Q^{\text{eff}}}(\theta_w, C_w)$ are the values used in subsequent modeling. Guarracino and Jougnot (2018) and Soldi et al. (2019) proposed analytical solutions to obtain these parameters, but with overall similar results.

A common form of the empirical relationship between the voltage coupling coefficient $C_{ek}^{sat}$ (mV/m) at saturation and electrical conductivity of the pore water $\sigma_w$ (S/m) is (Revil et al., 2003; Linde et al., 2007b; Jougnot et al., 2015):

$$\log_{10}|C_{ek}^{sat}| = a + b\log_{10}\sigma_w + (c\log_{10}\sigma_w)^2, \tag{8}$$

where Jougnot et al. (2015) found that $a = -0.6$, $b = -1.319$, $c = -0.1227$ are appropriate for modeling SP profiles at the HOBE hydrological site. From this, $\widehat{Q^{eff,sat}}(C_w)$ needed in Eq. (7) is obtained as (Revil and Leroy, 2004):

$$\widehat{Q^{eff,sat}} = -\frac{C_{ek}^{sat}(V/m)\sigma\rho_w g}{K_{sat}}, \tag{9}$$

where electrical conductivity $\sigma$ (S/m) is computed by Eq. (4), $\rho_w$ (kg/m$^3$) is the density of groundwater, $g$ (m/s$^2$) is the acceleration of gravity and $K_{sat}$ (m/s) is the saturated hydraulic conductivity.

*2.1.2. Electro-diffusive contribution*

Differences in ion mobility $\beta_j$ lead to electro-diffusive currents in the presence of electrochemical potential gradients. Considering Na$^+$ and Cl$^-$ as the only ions ($C_w = C_{Na^+} = C_{Cl^-}$) in the pore water, the electro-diffusive current density is written as (Pride, 1994; Revil and Jougnot, 2008; Jougnot et al., 2015):

$$\mathbf{J}_{ed} = -\frac{k_B T_K}{e_0}(t_{Na}^H - t_{Cl}^H)\sigma(\theta_w)\nabla \ln(C_w), \qquad (11)$$

where $T_K$ is the temperature in Kelvin (K), $e_0 = 1.602 \times 10^{-19}$ (C) is the elementary charge and $k_B = 1.3806 \times 10^{-23}$ (J/K) is the Boltzmann constant. Furthermore, $t_{Na}^H$ and $t_{Cl}^H$ are the microscopic Hittorf transport numbers related to the ion mobility $\beta_{Na}$ and $\beta_{Cl}$:

$$t_{Na}^H = \frac{\beta_{Na}}{\beta_{Na} + \beta_{Cl}}, t_{Cl}^H = 1 - t_{Na}^H. \qquad (12)$$

The microscopic Hittorf transport number $t_{Na}^H$ depends on concentration at high salinity, while it is constant at low salinities (Revil, 1999; Gulamali et al., 2011; Leinov and Jackson, 2014):

$$t_{Na}^H = \begin{cases} 0.3962, C_w < 0.09 \text{ mol/L} \\ 0.3655 - 9.2 \times 10^{-3} \ln(C_w), C_w \geq 0.09 \text{ mol/L} \end{cases}. \qquad (13)$$

When considering microporous media (e.g., clay), macroscopic Hittorf transport numbers (Revil and Jougnot, 2008) should be used to properly account for the anion exclusion effect (Leinov and Jackson, 2014). At the HOBE site, according to the investigation of grain-size distribution (Uglebjerg, 2013), the soil mostly consists of sand (> 90%), and the clay barely occurs in the shallow 50 cm (< 1%) and in the range of 100-250 cm depth (0.5% to 2%). We can, therefore, rely on the microscopic Hittorf transport numbers assuming a negligible cation exchange capacity (CEC) (see Jougnot et al. 2015 for comparative discussion of alternative formulations).

*2.1.3. Electrode effects*

Non-polarizable electrodes are electrodes for which the electrode polarization does not depend on the current flowing through the electrode. At the HOBE site, we rely on Petiau electrodes (Pb/PbCl$_2$, NaCl) that are described in detail by Petiau (2000) and manufactured by SDEC (PMS9000). These are electrodes of the second kind, which implies that an auxiliary salt (NaCl or KCl) is used in addition to the principal salt (PbCl$_2$). Petiau electrodes have been designed to ensure long-term stability (the electrode drift is on the order of 0.2 mV/month), minimal noise and low temperature dependence. The choice of Pb/PbCl$_2$ as a metal-salt couple is motivated by an extensive comparison with other alternatives by Petiau and Dupis (1980). The stability of Petiau electrodes is partly due to an appropriate combination of pH, and high principal salt (PbCl$_2$) and auxiliary salt (here NaCl, with KCl being a popular alternative) concentrations. Furthermore, the outlet in the form of a porous plug is located comparatively far (approximately 0.15 m) from the lead wire, and the design includes an internal thin channel located behind the porous plug enabling electrolytic contact while decreasing ionic diffusion. To further decrease ionic diffusion, the electrolyte is mixed with clay mud. With this design, the desaturation time, defined as the time for the diluted front originating at the outlet to reach the lead wire, is expected to be on the order of 14 years, assuming saturated conditions. Experiments mentioned by Petiau (2000) suggests that this time might be further increased by 20-40% when electrodes are installed under partially saturated conditions. This suggests that we can assume that the ionic concentrations of the principal and auxiliary salts in the vicinity of the lead wire are constant over many years. Nevertheless, this does not imply that the potential difference between the lead wire and the pore water outside the porous plug is constant over time.

A standard consideration is that the electrode potential depends on temperature, which implies that temperature differences between electrodes lead to unwanted voltages that need to be removed. For small temperature ranges, the self-potential due to temperature differences can be expressed by a linear relationship (Petiau, 2000):

$$SP_{\text{temp}} = k_{\text{T}}(T_{\text{obs}} - T_{\text{ref}}), \tag{14}$$

where the temperature coefficient $k_{\text{T}}$ depends on the electrode type. For Pb/PbCl$_2$ Petiau electrodes (Petiau, 2000) with a saturated NaCl electrolyte, $k_{\text{T}}$ is 0.2 mV/°C, $T_{\text{obs}}$ is the temperature in the position of the measurement electrode and $T_{\text{ref}}$ is the temperature at the reference electrode. To predict the temperature at any depth and time, we use the measured temperature data to fit a spatial-temporal distribution of temperature with a sine function:

$$T(t,z) = A(z)\sin\{\omega[t - \varphi_0(z)]\} + T_{\text{c}}, \tag{15}$$

where $\omega = \frac{2\pi}{365}$ (rad/d) is the angular frequency, $T_c$ is the annual average temperature that is assumed constant. The amplitude factor $A(z)$ and the phase delay $\varphi_0(z)$ is considered to vary linearly with depth. To further improve the corrections particularly at shallow depths, we add the interpolated residual values $res(t, z)$ as:

$$\tilde{T}(t,z) = T(t,z) + res(t,z). \tag{16}$$

A second non-standard consideration is that the membrane potential between the interior of the clay mud at concentration $C^{\text{ele}}$ and the pore water outside of the porous pot ($C^{\text{obs}}$ and $C^{\text{ref}}$ for the concentrations in the vicinity of the observation and reference electrodes, respectively) depends on the pore water chemistry. For practical purposes, we are interested in the differential membrane potential $SP_{\text{mb}}$, that is, the difference in membrane potential (Revil, 1999) between a measuring and reference electrode:

$$\begin{aligned} SP_{\text{mb}} &= -\frac{k_B T_K}{e_0}(2T^*_{\text{Na}} - 1)\ln\frac{C^{\text{ele}}}{C^{\text{obs}}} - \left[-\frac{k_B T_K}{e_0}(2T^*_{\text{Na}} - 1)\ln\frac{C^{\text{ele}}}{C^{\text{ref}}}\right] \\ &= \frac{k_B T_K}{e_0}(1 - 2T^*_{\text{Na}})\ln\frac{C^{\text{ref}}}{C^{\text{obs}}}, \end{aligned} \tag{17}$$

where $T^*_{\text{Na}}$ is the macroscopic Hittorf transport number in the clay mud filling the Petiau electrodes. Determining this value is difficult as the detailed design of the mud mixture in the PMS9000 electrodes is unknown. Using Petiau (2000) as a guide, we make the following assumptions concerning the clay mud in the electrodes: cation exchange capacity CEC = 2.89× $10^3$ C/kg (0.03 in meq/g), density $\rho_e$ = 2600 kg/m³, porosity $\phi_e = 0.5$ and cementation exponent $m_e = 1.5$. With these values, the total excess of charge $Q_v$ can be calculated by (Waxman and Smits, 1968):

$$Q_v = \rho_e \frac{1-\phi_e}{\phi_e} \text{CEC}. \tag{18}$$

This volumetric excess charge $Q_v$ of the clay mud is needed to determine the surface electrical conductivity $\sigma_s$ and is different in nature than the flux-averaged effective excess charge density described in Eq. (5) (see discussions in Jougnot et al., 2019; Jougnot et al., 2020). The ratio of $\sigma_s$ to the electrical conductivity of electrolyte $\sigma_w$ is (Sen and Kan, 1987; Revil et al., 1998):

$$\gamma_e = \frac{\sigma_s}{\sigma_w} \approx \frac{2\beta_s Q_v}{3\sigma_w}\left(\frac{\phi_e}{1-\phi_e}\right) = \frac{2\beta_s \rho_e \text{CEC}}{3\sigma_w}, \tag{19}$$

where $\beta_s$ is the clay surface ionic mobility taken as $5.14 \times 10^{-9}$ (m$^2$s$^{-1}$V$^{-1}$) for Na$^+$ (Revil, 1999). The macroscopic Hittorf number $T_{\text{Na}}^*$ grows as a function of this ratio $\gamma_e$ following (Revil, 1999):

$$T_{\text{Na}}^* = \left[1 + \frac{1-t_{\text{Na}}^*}{\gamma_e \phi_e^{-m_e} + \frac{1}{2}(t_{\text{Na}}^* - \gamma_e)\left(1 - \frac{\gamma_e}{t_{\text{Na}}^*} + \sqrt{\left(1-\frac{\gamma_e}{t_{\text{Na}}^*}\right)^2 + \frac{4\gamma_e \phi_e^{-m_e}}{t_{\text{Na}}^*}}\right)}\right]^{-1}, \tag{20}$$

where the microscopic Hittorf number $t_{\text{Na}}^*$ of Na$^+$ in the clay mixture of the electrode can be estimated by Eq. (13) assuming that $C_e$ is 6.8 mol/L (Petiau, 2000). Based on the above assumptions, $T_{Na}^*$ of the electrode clay mixture is 0.3492, which is only slightly larger than $t_{Na}^*$ due to the high concentration.

*2.1.4. Total SP*

The electrical potential response of the streaming current $\mathbf{J}_{ek}$ and the electro-diffusive source $\mathbf{J}_{ed}$ are computed and indicated as $SP_{ek}$ and $SP_{ed}$, respectively. The total SP signal is:

$$SP = SP_{ek} + SP_{ed} + SP_{temp} + SP_{mb}. \tag{21}$$

We design $SP_{sim} = SP_{ek} + SP_{ed} + SP_{mb}$ as the simulated SP response that is to be compared with the temperature-corrected raw data using Eqs. (14) and (15) that we refer to as $SP_{obs}$. A modified version of the finite difference MaFlot code (Künze and Lunati, 2012) is used to simulate the SP response by solving the Poisson's equation obtained by combining Eqs. (1) and (2). Figure 1a displays a sketch to illustrate the calculation of total SP.

*2.2. Hydrodynamic characterization*

To apply the framework of section 2.1, we need inputs in terms of distributed time-series of water content, water fluxes and water concentrations. To achieve this, we use a finite element solution of the Richards equation and the advection-dispersion equation.

*2.2.1. Water flow and solute transport*

Richards equation is used to simulate water flow under variably-saturated conditions (Richards, 1931):

$$\frac{\partial \theta_w}{\partial t} + S = \nabla(K \nabla H_p), \quad (22)$$

where $S$ (s$^{-1}$) is the source term, $H_p$ (m) is the water pressure head, and $K$ (m/s) is the hydraulic conductivity expressed as:

$$K(\theta_w) = K_{rel}(\theta_w) K_{sat}, \quad (23)$$

where $K_{rel}$ (-) is the relative hydraulic conductivity described by the van Genuchten (1980) model with parameters $\alpha_{VG}$ and $n_{VG}$.

For given boundary and initial conditions, $H_p$ and $\theta_w$ can be calculated by the governing equation Eq. (22). The resulting water flux $\mathbf{u}$ is expressed as:

$$\mathbf{u} = -K(\theta_w) \nabla H_p. \quad (24)$$

For solute transport, we rely on the advection-dispersion equation (ADE):

$$\frac{\partial(\theta_w C_j)}{\partial t} + \nabla \cdot [\mathbf{u} C_j - \theta_w (D_j + \alpha \bar{v}) \nabla C_j] = 0, \quad (25)$$

where $\bar{v} = \frac{u}{\theta_w}$ is the average flow velocity, $D_j$ (m²/s) is the ionic diffusion coefficient and $\alpha$ (m) is the dispersivity (Bear, 2012).

Using a conservative ADE, the fresh infiltrating rainwater will in a recharge-dominated setting eventually fill the saturated portion of the pore space. This is unrealistic as demonstrated by the fact that the electrical salinity of the undisturbed pore water is around 0.02 S/m (Jougnot et al., 2015). Coupling the flow- and transport simulations with a predictive soil chemical model including interactions between the soil minerals and the water phase is extremely challenging and outside the scope of this study. In an effort, to account for the leading effect of how fresh rainwater is turned into more charged pore water, we use a very simple iterative approach (Fig. 1a). We output the profile of simulated concentrations $C_{cur}$ daily ($dt = 1$ day) and compare with an assumed equilibrium concentration $C_{eq}$:

if $C_{cur} < C_{eq}$ then $C_{in} = \gamma dt(C_{eq} - C_{cur}) + C_{cur}$, (26)

else $C_{in} = C_{cur}$,

where $\gamma$ determines how quickly equilibrium is reached. Considering two extreme conditions: if $\gamma = 0$ d⁻¹, no adaptation is made and $C_{in} = C_{cur}$; if $\gamma = 1$ d⁻¹, we have $C_{in} = C_{eq}$. The resulting $C_{in}$ is used as the initial concentration when simulating the following day. We stress that this adaptation in the modeling framework is made to qualitatively assess the impact of rain water properties on SP phenomena in the near surface and that it is not a consistent physics-based model.

*2.2.2. Inversion of hydrodynamic parameters*

Based on Linde et al. (2006), we express the effective dielectric constant $\kappa_t^{\text{sim}}$ as

$$\kappa_t^{\text{sim}} = \frac{(F-1)\kappa_s + \left(\frac{\theta_w}{\phi}\right)^n \kappa_w + \left[1 - \left(\frac{\theta_w}{\phi}\right)^n\right]\kappa_a}{F}, \qquad (27)$$

where $\kappa_s = 4.3$ and $\kappa_a = 1$ are the dielectric constants of the solid grains and air, respectively. $\kappa_w$ is the dielectric constant of pore water, which depends on temperature $T$ (°C) following (Weast et al., 1988; Wraith and Or, 1999):

$$\kappa_w = 78.54[1 - 4.579 \times 10^{-3}(T - 25) + 1.19 \times 10^{-5}(T - 25)^2 - 2.8 \times 10^{-8}(T - 25)^3]. \qquad (28)$$

The observed temperature data at the field site is used in Eq. (28) to estimate the dielectric constant of water $\kappa_w$ before using Eq. (27) to obtain the simulated effective dielectric constant $\kappa_t^{\text{sim}}$ from the simulated water contents. We contrast the $\kappa_t^{\text{sim}}$ with the observed effective dielectric constant $\kappa_t^{\text{obs}}$ to evalute the likelihood of a given vadose zone model by using an $\ell_1$-norm of residual errors. This leads to the following Laplacian log-likelihood function:

$$\mathcal{L}\left(\theta_w^r, \phi, \alpha_{\text{VG}}, n_{\text{VG}}, K_{\text{sat}} | \kappa^{\text{obs}}\right) = -\sum_{t=1}^{N}\sum_{i=1}^{M}\left(\ln(2\sigma_t^i)\right) - \qquad (29)$$

$$\sum_{t=1}^{N}\sum_{i=1}^{M}\left(\frac{\left|\kappa_t^{\text{obs},i}-\kappa_t^{\text{sim},i}\right|}{\sigma_t^i}\right),$$

where $\sigma_t^i$ is a data error, $M$ is the total number of available sensors and $N$ is the total number of times considered. The log-prior density is also taken into account when calculating the log-posterior density of each model realization.

$$p(\theta_w^r, \phi, \alpha_{\text{VG}}, n_{\text{VG}}, K_{\text{sat}}|\kappa^{\text{obs}}) = \mathcal{L}(\theta_w^r, \phi, \alpha_{VG}, n_{VG}, K_{\text{sat}}|\kappa^{\text{obs}}) + \qquad (30)$$
$$\sum_{i=1}^{q}\sum_{k=1}^{5}\ln[\mathbb{N}(\mu_i^k, \sigma_i^k, X_i^k)],$$

where for $q$ layers $\mathbb{N}(\mu_i^k, \sigma_i^k, X_i^k)$ represents the normally-distributed prior distribution with $\mu_i^k$ and $\sigma_i^k$ the expected value and standard deviation of parameter $k$ at the $i$th layer, $X_i^k$ represents the sampled parameter value of $\theta_w^r$, $\phi$, $\alpha_{\text{VG}}$, $n_{\text{VG}}$ and $K_{\text{sat}}$ respectively.

The inversion process follows the diagram in Fig. 1b. As indicated by Fig. 1b, we apply Markov chain Monte Carlo (MCMC) to infer the maximum a posteriori estimate of the model parameters. For this, we rely on the DREAM$_{\text{(ZS)}}$ algorithm (Laloy and Vrugt, 2012; Vrugt, 2016). We refer to Jougnot et al. (2015) for more details.

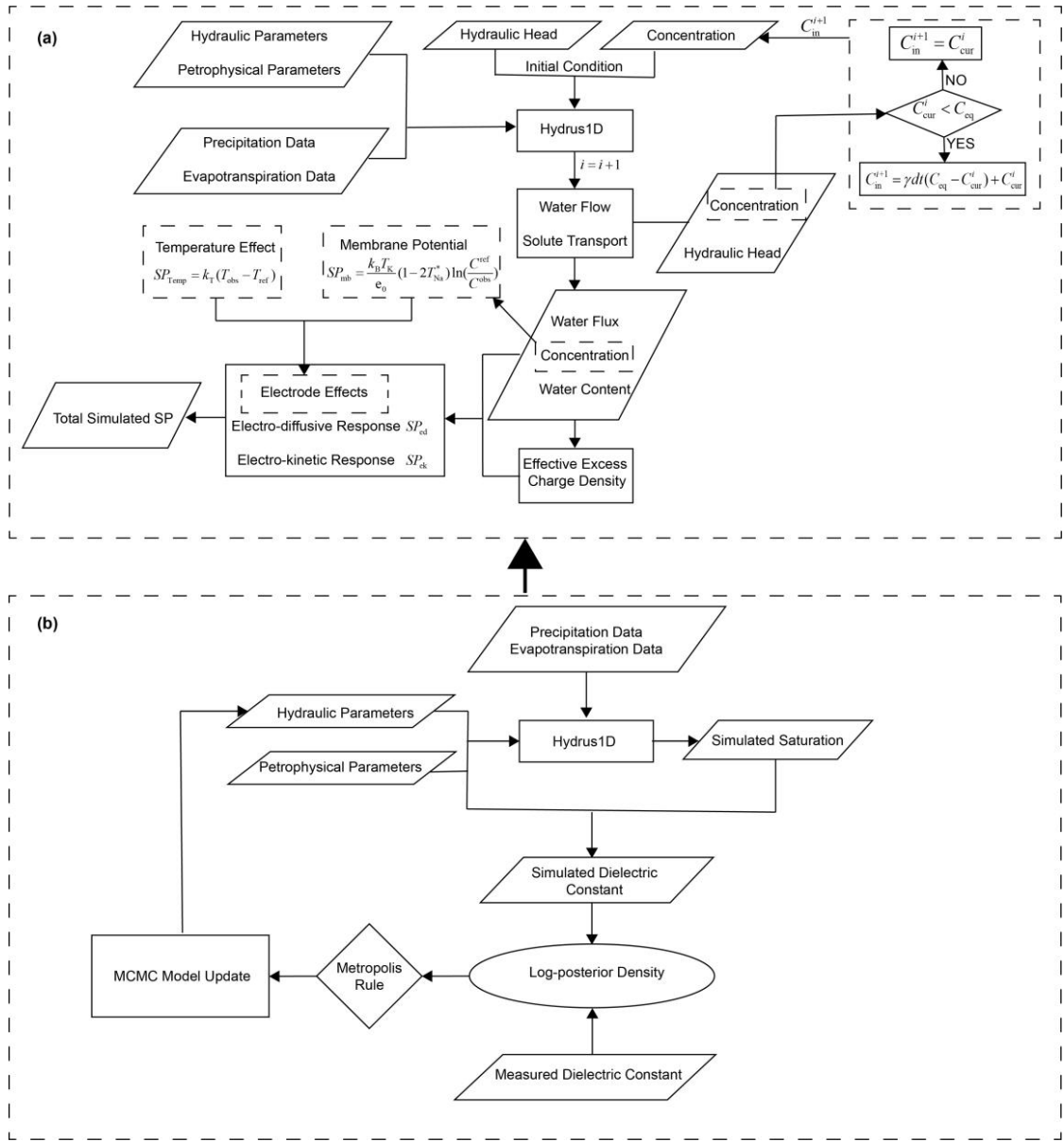

Fig. 1. Flow diagrams for simulating SP signals (modified from Jougnot et al., 2015) displaying (a) the SP simulation scheme and (b) the inversion scheme for hydraulic parameters.

## 3. The HOBE observatory and data

The HOBE hydrological observatory is located within an agricultural field (Fig. 2) in the Skjern river catchment (Fig. 2a), near Voulund, western Denmark (Jensen and Illangasekare, 2011; Jensen and Refsgaard, 2018). It was established to study hydrological responses generated by natural and artificial forcings at different scales and to establish water balance closure in the context of assessing and managing water resources within the catchment by integration of data from multiple measuring techniques (e.g., geophysics, hydrology, meteorology, remote sensing, geochemistry). The patterns of dynamic hydrological landsurface and subsurface processes are analyzed based on multi-scale monitoring in space and time. Previous findings relevant to this work include (1) a description of the geological architecture and soil components based on analysis of drilling samples (Uglebjerg, 2013); (2) the estimation of potential and actual evapotranspiration (Ringgaard et al., 2011) and the establishment of the water balance at the field-scale (Vásquez et al., 2015); (3) the detection of soil moisture and other pools of hydrogen using the cosmic-ray neutron method (Andreasen et al., 2016; Andreasen et al., 2017); (4) monitoring the pathway of saline movement based on cross-borehole electrical resistivity tomography (ERT) and ground-penetrating radar (GPR) (Haarder et al., 2015) and from SP data (Jougnot et al., 2015).

Based on the analyzed samples from five drill cores (Uglebjerg, 2013), the soil is described by seven horizontal layers. In each layer, more than 90% of the soil is made up of very fine to very coarse sand, with silt and clay content in layers 1 and 2 being 5-10%. In the present work, we assume that surface conductivity is negligible given the soil description and that we did not have access to induced polarisation data (e.g., Revil et al., 2019) at the site. The water table is fluctuating between 5.5 m and 6.5 m depth. In September 14, 2011, 40 liters of saline water with an electrical conductivity of 22.1-24 S/m was injected during two hours uniformly over a 12 m × 12 m experimental plot. The tracer experiment was described and analyzed by Haarder et al. (2015) and the SP signature of the saline injection was studied by Jougnot et al. (2015).

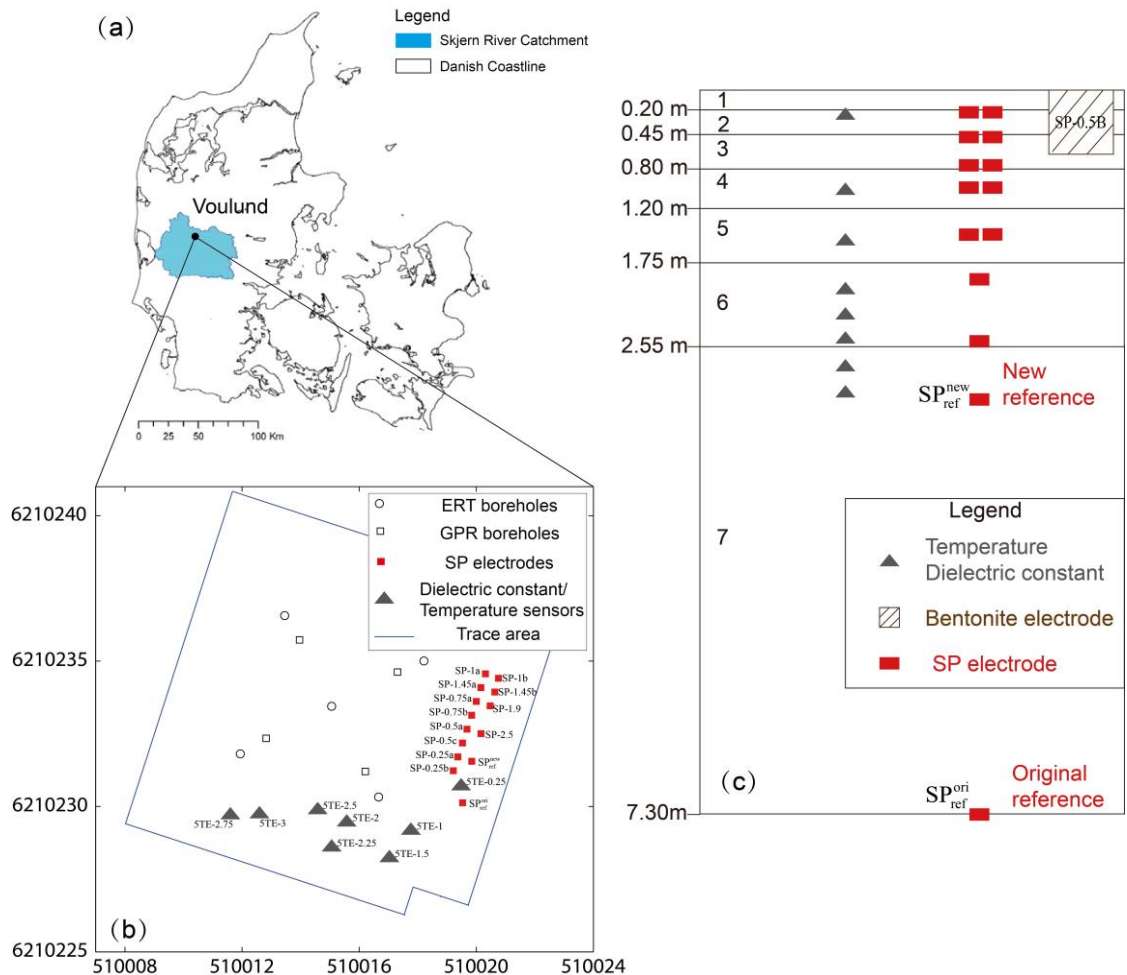

Fig. 2. (a) Map of Denmark with (b) top view with locations of sensors and (c) depth information about the sensors (modified from Jougnot et al., 2015). Electrode SP-0.5B was installed several meters away from the area depicted in (b).

*3.1. Meteorological data*

Precipitation data have been collected hourly at the site since April 2009 using multiple rain gauges. A time series was developed, combining data from five rain gauges, in order to obtain what was named *the Best Assembled Precipitation Dataset*. In this dataset, the majority of the data (~70%) came from a ground-based rain gauge located within 25 m from the SP installations (see below). Hourly potential evapotranspiration data have been calculated with the Penman-Monteith equation (Monteith, 1965) using wind speed, temperature, humidity, solar radiation and crop height data from November 2008 until August 2013. Based on the analysis of rain water, Uglebjerg (2013) found that its average electrical conductivity is approximately 50 µS/m.

*3.2. Temperature and dielectric constant data*

Temperature, dielectric constant measured at 70 MHz and electrical conductivity data have been recorded every 20 minutes using 17 Decagon 5TE probes (www.decagon.com) distributed in the upper 3 m of the soil since July 2011. Water content was estimated using temperature and dielectric constant data (see section 2.2). During the long-term monitoring, several of the sensors became dysfunctional, especially the electrical conductivity sensors. In this study, we rely on eight sensors of temperature and dielectric constant (shown by grey triangles in Fig. 2c) that acquired data during the full considered time-period. They are located at depths of 0.25 m, 1 m, 1.5 m, 2 m, 2.25 m, 2.5 m, 2.75 m and 3 m, repectively.

*3.3. Calibration of hydrological parameters*

In order to predict the SP response, we first use a similar approach (see section 2.2.2) as Jougnot et al. (2015) to calibrate the hydraulic parameters, with the main difference that we rely on a longer time-record (December 1, 2011 to July 22, 2013) of dielectric constant and temperature data. We kept the petrophysical parameters $m$ and $n$ values in Eq. (27) to those obtained by Jougnot et al. (2015), namely 1.38 and 1.57, respectively. For each tested combination of soil parameter values, Hydrus-1D (Šimůnek et al., 2013) was used to calculate the water content $\theta_w$ at the sensor locations (0.25 m, 1 m, 1.5 m, 2 m, 2.25 m, 2.5 m, 2.75 m, and 3 m). We initialize our simulations on August 1, 2010 assuming a pressure head profile of -50 cm. The time-varying flux boundary condition at the surface is determined by precipitation and potential evapotranspiration data, except for the period of saline injection in September 2011 during which the vertical inflow was imposed at 3.6 cm/d. At the bottom of the numerical domain with thickness of 6.5 m, a fixed pressure head of 50 cm is used corresponding to an assumed constant water table at 6 m. The profile is divided into 7 soil layers that are discretized using 300 nodes. From the flow simulations, we output the water content $\theta_w$ daily at midnight in the period from December 1, 2011 to July 22, 2013.

**Table 1:** Uncorrelated prior Gaussian distribution of model parameters (modified from Jougnot et al., 2015) with $\mu$ and $\sigma$ indicating the mean value and the standard deviation, respectively. $K_{sat}$ is saturated hydraulic conductivity and $\theta_w^r$ is residual water content.

| Layer | Depth (m) | $\theta_w^r$ | | $\phi$ | | $\log_{10} \alpha_{VG}$ (cm$^{-1}$) | | $n_{VG}$ | | $\log_{10} K_{sat}$ (cm/d) | |
|---|---|---|---|---|---|---|---|---|---|---|---|
| | | $\mu$ | $\sigma$ | $\mu$ | $\sigma$ | $\mu$ | $\sigma$ | $\mu$ | $\sigma$ | $\mu$ | $\sigma$ |
| 1 | 0-0.2 | 0.05 | 0.01 | 0.39 | 0.01 | -0.87 | 0.5 | 1.36 | 0.1 | 3.31 | 1.47 |
| 2 | 0.2-0.45 | 0.04 | 0.01 | 0.38 | 0.02 | -1.15 | 0.32 | 2.3 | 0.36 | 2.74 | 0.49 |
| 3 | 0.45-0.8 | 0.03 | 0.01 | 0.38 | 0.02 | -1.29 | 0.1 | 2.3 | 0.36 | 2.44 | 0.46 |
| 4 | 0.8-1.2 | 0.07 | 0.03 | 0.4 | 0.02 | -1.02 | 0.17 | 1.71 | 0.31 | 2.13 | 0.12 |
| 5 | 1.2-1.75 | 0.06 | 0.02 | 0.37 | 0.04 | -0.96 | 0.14 | 2.49 | 0.64 | 2.46 | 0.43 |
| 6 | 1.75-2.55 | 0.04 | 0.01 | 0.39 | 0.02 | -1.21 | 0.36 | 2.89 | 1.04 | 2.46 | 0.43 |
| 7 | 2.55-7.3 | 0.04 | 0.01 | 0.39 | 0.02 | -1.29 | 0.22 | 2.89 | 1.04 | 2.42 | 0.43 |

Gaussian prior distributions of material properties were used (Table 1) and the initial models were obtained as random draws from the prior. The Metropolis acceptance rule (Metropolis et al., 1953) is used to determine, at each MCMC step, if the proposal is accepted. We use three MCMC chains each having a length of 2000. The number of layers $q$ is 7 in Eq. (30). The standard deviation of the dielectric constant data, $\sigma_t^i$, appearing in Eq. (29) data is assumed constant and equal to 1. This value is not so crucial as we are interested in the maximum a posteriori estimate; it was determined by trial-and-error and describes how well we are able to fit our observed data. The log-likelihood value increases rapidly during the first 200 MCMC steps before increasing more gradually. Using Eq. (30), we calculate the corresponding log-posterior densities and the maximum value is picked. The corresponding model parameter values are then used in the subsequent SP simulations (Table 2). Figure 3 is a scatter plot comparing the simulated and measured dielectric constant data for this best-fitting model. The average absolute mean deviation is 0.7168. The simulated dielectric constant underestimates the variability in the observed data, particularly during the winter season when the water content can be very high due to decreased evapotranspiration and snow melt events (Haarder et al., 2015).

**Table 2:** The parameter values of the maximum a posteriori model obtained by MCMC inversion.

| Layer | Depth (m) | $\theta_w^r$ | $\phi$ | $\alpha_{VG}$ (cm$^{-1}$) | $n_{VG}$ | $K_{sat}$ (cm/d) |
|---|---|---|---|---|---|---|
| 1 | 0-0.2 | 0.066 | 0.40 | 0.025 | 1.375 | 1273 |
| 2 | 0.2-0.45 | 0.071 | 0.34 | 0.032 | 1.935 | 197.4 |
| 3 | 0.45-0.8 | 0.023 | 0.42 | 0.045 | 2.842 | 39.97 |
| 4 | 0.8-1.2 | 0.020 | 0.39 | 0.054 | 1.760 | 300.6 |
| 5 | 1.2-1.75 | 0.064 | 0.32 | 0.092 | 2.031 | 829.2 |
| 6 | 1.75-2.55 | 0.069 | 0.37 | 0.045 | 2.240 | 589.9 |
| 7 | 2.55-7.3 | 0.053 | 0.32 | 0.043 | 3.454 | 38.80 |

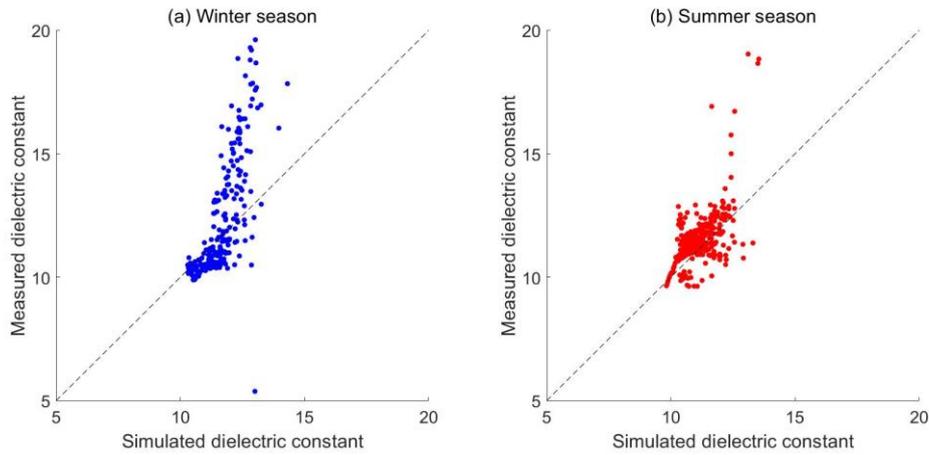

Fig. 3. Scatter plot of simulated dielectric constant against measured dielectric constant in the (a) winter season (October to March) and the (b) summer season (April to September) of 2012.

*3.4. Self-potential data*

With assuming only vertical flow, a total of 15 non-polarizable Petiau electrodes ($Pb/PbCl_2$, $NaCl$) were installed at depths of 0.25 m, 0.5 m, 0.75 m, 1 m, 1.45 m, 1.9 m, 3.1 m and 7.3 m in individual holes (Fig. 2c). Theses electrodes holes were drilled by augering. Each electrode was inserted vertically in a thin hole and it was attached to a wooden stick to ensure that the electrode had a good contact with the bottom of the hole and to better estimate the depth of installation. The one exception to this procedure was one of the top electrodes that was installed at 0.5 m in an individual dug hole, which was later filled with bentonite. As shown in Fig. 2c, there are two electrodes installed without bentonite surroundings at 0.25 m, 0.5 m, 0.75 m, 1 m and 1.45 m. The reference electrode during the measurements was located well below the water table at 7.3 m depth. To remove the strong influence of water table fluctuations from the acquired SP data, they were processed to have the deepest electrode in the vadose zone (at 3.1 m depth) serve as the reference electrode. Due to the superposition principle, this is achieved by subtracting the measured SP data at this location from all the other measured SP time-series. Initially, from July 19, 2011 until September 9, 2011, SP data were recorded every 20 minutes. Subsequently, SP data were recorded every 5 minutes. Herein, we present data until July 28, 2015. Repeated crosshole ERT measurements were performed in the area from September 2011 to August 2012 (Haarder et al., 2015), leading to large recorded voltages that are unrelated to the processes of interest. To filter out these effects, we first median filter the SP data at each recorded time $t_i$ within a period of $t_i - 75$ minutes to $t_i + 75$ minutes. Second, in the period of ERT measurements we identify periods $\Delta t_k$ disturbed by the ERT measurements by identifying times with anomalously high temporal SP variations. For these time-periods, we replace the raw data with the median filtered data. Finally, we median filter the resulting time-series using a 1-hour window corresponding to the time-interval used to represent the SP data in this paper. See Mariethoz et al. (2015) for an alternative data processing approach that has been demonstrated for a sub-set of the presented data.

## 4. Results

*4.1. Calibration and sensitivity analysis of electrokinetic and electrochemical effects*

We first consider the streaming potential generated by electrokinetic effects. In doing so, we focus on the SP data acquired prior to the saline tracer injection. In this period immediately following the installation, the electrodes are expected to be stable and the electro-diffusive effect should be minimal. In addition to simulations using the WR and RP methods mentioned in section 2.1.1, we also consider modeling based on the thick double-layer assumption in which the effective excess charge density is scaled with water saturation $\frac{\phi}{\theta_w}$ (Linde et al., 2007a):

$$\widehat{Q^{\text{eff}}_{\text{Linde}}}(\theta_w, C_w) = \frac{\phi}{\theta_w} \widehat{Q^{\text{eff,sat}}}(C_w), \tag{31}$$

where the saturated excess charge density $\widehat{Q^{\text{eff,sat}}}(C_w)$ is estimated by Eqs. (8) and (9). We present SP profiles (Figs. 4b-d) at three times corresponding to before, during and after a rainfall event (Fig. 4a). As expected for a recharge-dominated regime with associated downward flow, the recorded SP data (Figs. 4b-d) become increasingly negative towards the surface. Furthermore, the electrodes located at the same depth display similar SP values indicating that electrode-related effects are uniform. The simulated SP signals have similar shapes (Figs. 4b-d), but it is only the WR method that provides simulations of the same magnitude as the observed data. The RP method overestimates the amplitudes roughly by a factor of two and the thick double-layer assumption of Eq. (31) leads to simulated SP signals that are roughly 50% too small. The temporal differences in the observed SP data (Figs. 4e-g) display small decreases in SP magnitudes, which is in contrast to the simulations that show larger increases. In the following, we only present predictions based on the WR method. In the simulations described above, we used calibrated values of the inverse time constant $\gamma = 0.025 \text{ d}^{-1}$ (Eq. 26) obtained by maximizing the correlation between simulated and observed SP signals. As shown in Figs. 4h-j, the vertical SP profiles are only slightly affected by the delay factor for $\gamma > 0.025$ d$^{-1}$.

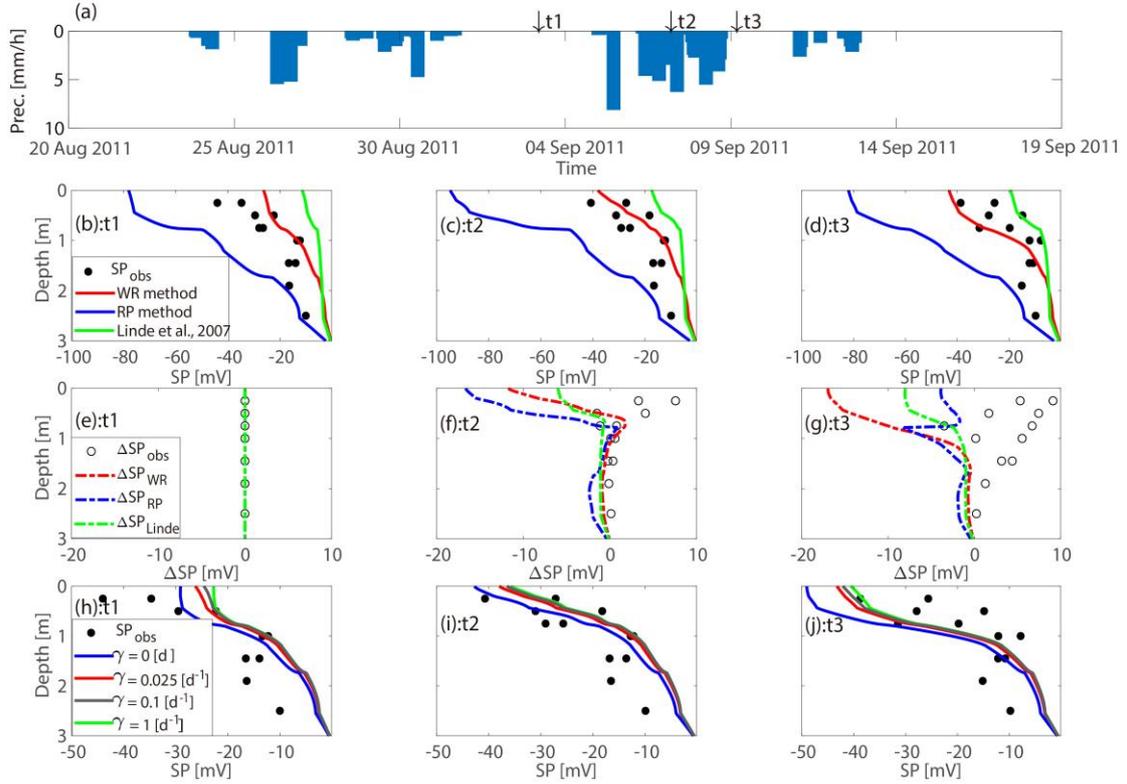

Fig. 4. (a) Precipitation data with three chosen times (t1-t3) highlighted by black arrows. Comparison between observed and simulated SP profiles (b) before a heavy rain at t1, (c) during the rain at t2 and (d) after the rain at t3. The solid lines are simulated SP data using different excess charge densities $\widehat{Q^{\text{eff}}}$ obtained, respectively, from the WR method (red), the RP method (blue) and the thick double-layer model (green) of Linde et al. (2007). (e-g) Differences of observed SP (circles) and simulated SP (dashed lines) with respect to t1. (h-j) Simulated SP signals (solid lines) using the WR method with different γ-values in equation (26).

After choosing the WR method for calculating the effective charge density (Eq. 7) and calibrating the delay factor at $\gamma = 0.025$ d$^{-1}$ (Eq. 26) using the observed SP data prior to saline injection, we now consider the SP simulations for longer time-series including the saline tracer injection period. Following the ERT-based estimations of the saline tracer test by Haarder et al. (2015), the dispersivity α (Eq. 25) is initially set to 0.25 m. We compare the SP simulations with the observed SP signals from July 2011 to September 2012 at 0.25 m and 0.75 m (Fig. 5). The simulated SP data given as a superposition of electrokinetic and electro-diffusive contributions without electrode-related effects are presented in Figs. 5b and 5e. The simulated SP signals describe the main shape of the observed SP data rather well, particularly the strong response caused by the saline tracer injection. To assess electrode-related effects, we first consider the estimated macroscopic Hittorf number $T_{Na}^*$ of the electrode clay-mixture calculated by Eq. (20). The simulated SP signals including the electrode-effect (Figs. 5c and 5f) display a much slower decrease in SP amplitudes after the tracer injection than what is observed, and the gradual relaxation observed in the data towards more negative values is not reproduced. We then sought the macroscopic Hittorf number $T_{Na}^*$ that maximises the correlation between the observed and simulated SP data. In Figs. 5d and 5g, we display the prediction for the best-fitting calibrated value of 0.43. This higher value is rather consistent with the fact that the clay allows a larger mobility of the cation due to an important volume of EDL in its pore space (see discussion in Revil and Jougnot, 2008). Here, the simulated responses are found to be intermediate between the two other cases. For the shallow electrode, we see that the decrease of SP magnitudes (Fig. 5d) following the saline tracer injection is sharper than for the other cases (Figs. 5c-f) and in better agreement with the observed data.

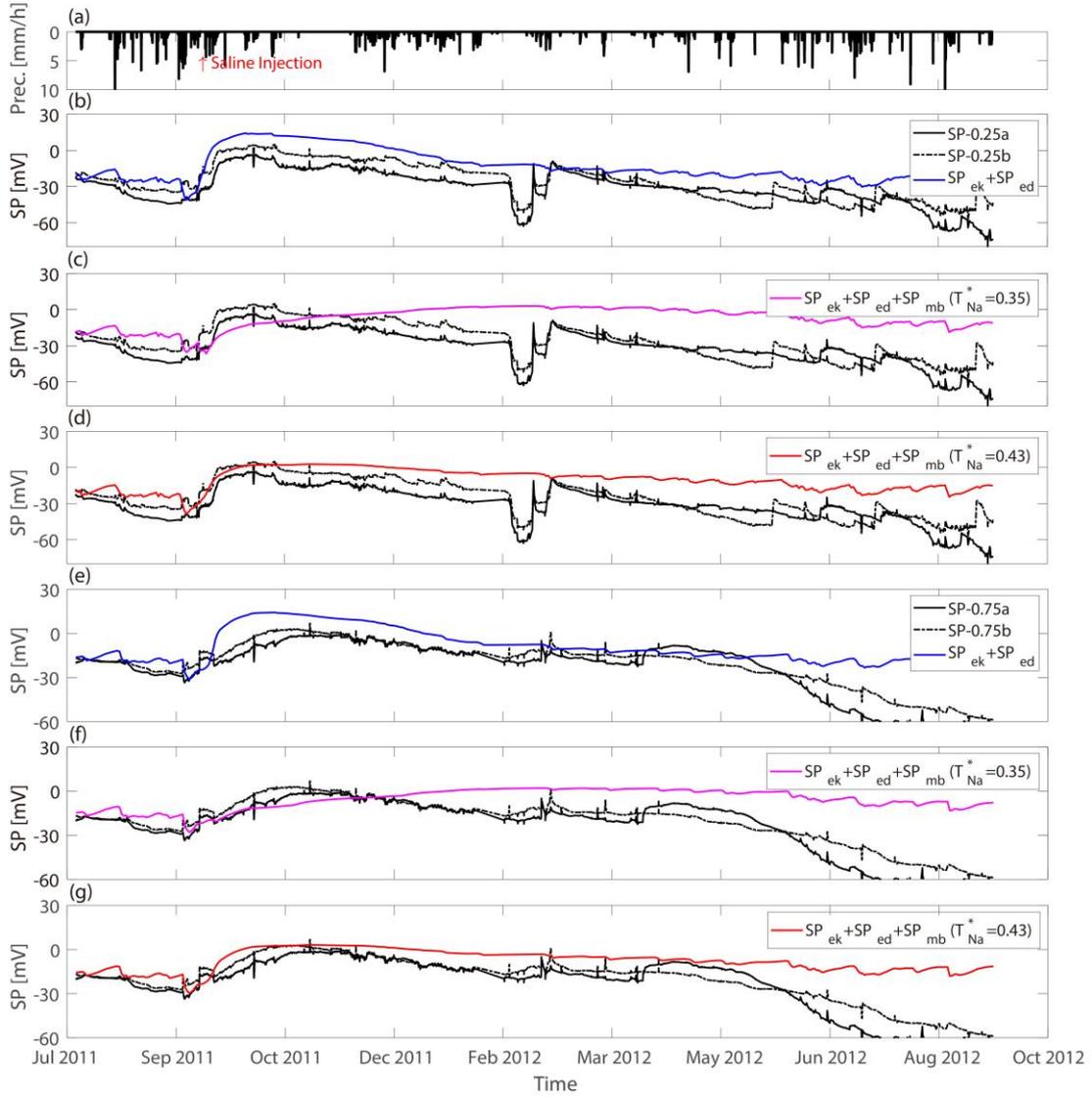

Fig.5. (a) Precipitation data during the first year of SP monitoring with the red arrow indicating the time of saline tracer injection. Solid black and dashed lines indicate the observed SP data at two measuring electrodes located at (b-d) 0.25 m and (e-g) 0.75 m depth. (b, e) The sum of the simulated streaming potential and electro-diffusive potential is shown in blue. The sum of the simulated streaming potential, electro-diffusive potential and electrode membrane potential with (c, f) the calculated macroscopic Hittorf number of the electrode clay mixture in purple and (d, g) with the best fitting macroscopic Hittorf number (0.43) in red.

Using the optimized $T_{Na}^{*} = 0.43$ for the electrode clay mixture, we now investigate the sensitivity of our simulated SP responses to the dispersivity α. The impact of α at 0.25 m (Fig. 6a) and 0.75 m (Fig. 6b) mainly concerns the speed at which the SP magnitudes decrease following the tracer injection, with faster decreases observed for larger α, and the rate of relaxation towards more negative values being the strongest for small α. The most satisfactory simulations are offered by α = 0.1 m and this value is retained in the following. As an aside, we note that the observed SP curves at 0.25 m (Fig. 6a) present a notable decrease in February 2012, which is attributed to the partially frozen topsoil (and electrode). Indeed, the soil temperature at this depth is below 0 ℃ in this time period.

In Fig. 7, we present the simulated spatio-temporal evolution of the solute concentration, which suggests that all of the injected saline should have reached the groundwater in April 2013. The trajectory of the simulated maximum concentration agrees well with the ERT-based estimates (Haarder et al., 2015), with both showing an acceleration of the plume movement in the winter. We note further that our estimates are in-between the ERT-based estimates of Haarder et al. (2015) and those determined by analyzing core data.

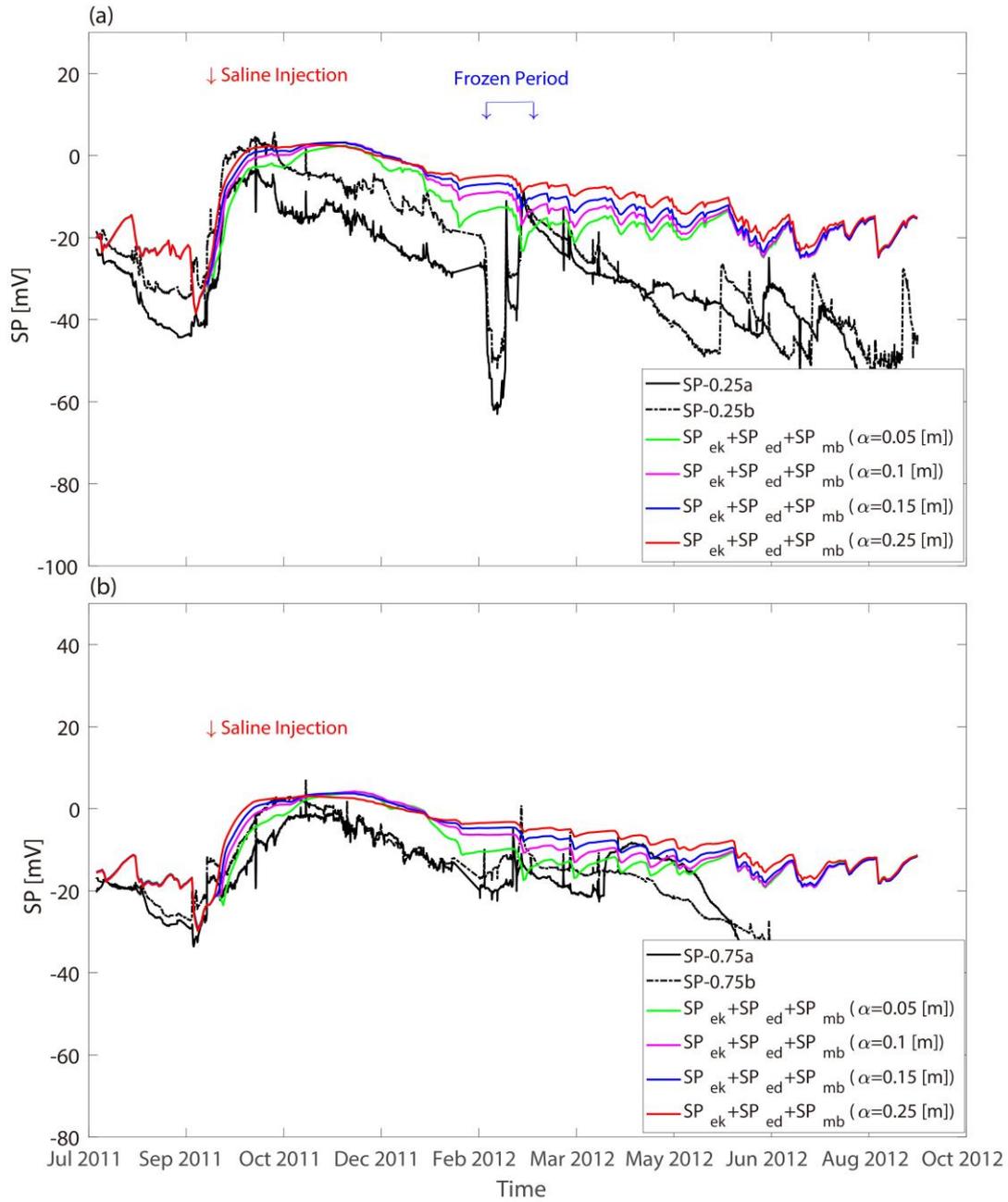

Fig. 6. Comparison between the observed (black solid and dashed lines) and simulated SP signals using the best fitting macroscopic Hittorf number (0.43) for different values of dispersivity at (a) 0.25 m and (b) 0.75 m depth. The chosen values of dispersivity are 0.05 m (green), 0.1 m (purple), 0.15 m (blue) and 0.25 m (red). The red arrow indicates the time of saline tracer injection and the blue arrows indicate the frozen period.

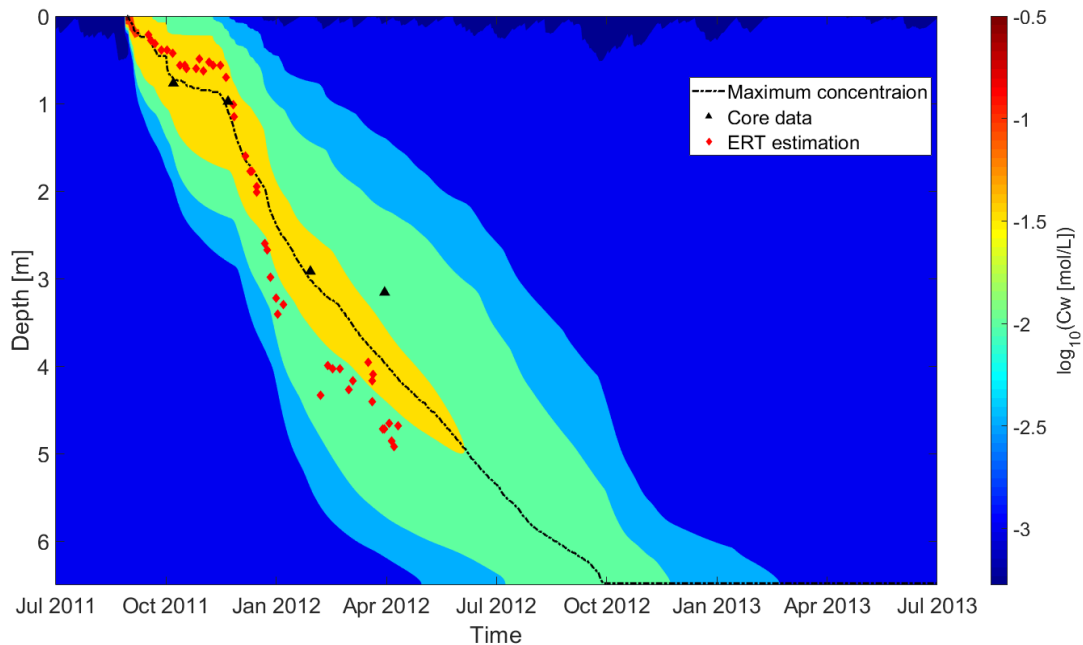

Fig. 7. The spatio-temporal variations of the simulated solute concentration of pore water assuming a dispersivity of 0.1 m. The dashed line corresponds to the depth of maximum concentration, the black triangles indicate the center of solute mass determined by drilled core data and the red diamonds are the center of solute mass inferred by Haarder et al. (2015) using ERT data.

*4.2. Electrode stability*

Using the calibrated parameters, we simulate the SP responses as a superposition of electrokinetic, electro-diffusive, and electrode-related effects over four years and consider five vertical SP profiles in the summer seasons of 2011, 2012, 2013, 2014, and 2015 (Fig. 8). The corresponding dates were chosen to correspond to periods with heavy rainfall implying that the simulated SP data have similar shapes and magnitudes at all of the considered dates. However, it is only in 2011 (Fig. 8a) shortly after the SP installations that the observed SP data also show a trend of increasingly negative values towards the surface as expected for an infiltration-dominated system. At later years, the observed data start to deviate from this expected behavior. The observed data are negative in 2012 (Fig. 8b), but the amplitudes are higher and there are rather significant oscillations with depth. This pattern is similar in 2013 (Fig. 8c), but with smaller magnitudes. The observed SP data turn positive in 2014 (Fig. 8d) and the magnitudes of the most positive values in 2015 approach 200 mV (Fig. 8e), thus, the different scale used on x-axis of Fig. 8e. Since no major changes in the state of the vadose zone is expected between these years as indicated by the simulated data, this suggests that the observed differences between the years are largely due to time-varying electrode polarizations manifesting themselves by electrode-potentials that develop differently with depth (different water content magnitudes and dynamics).

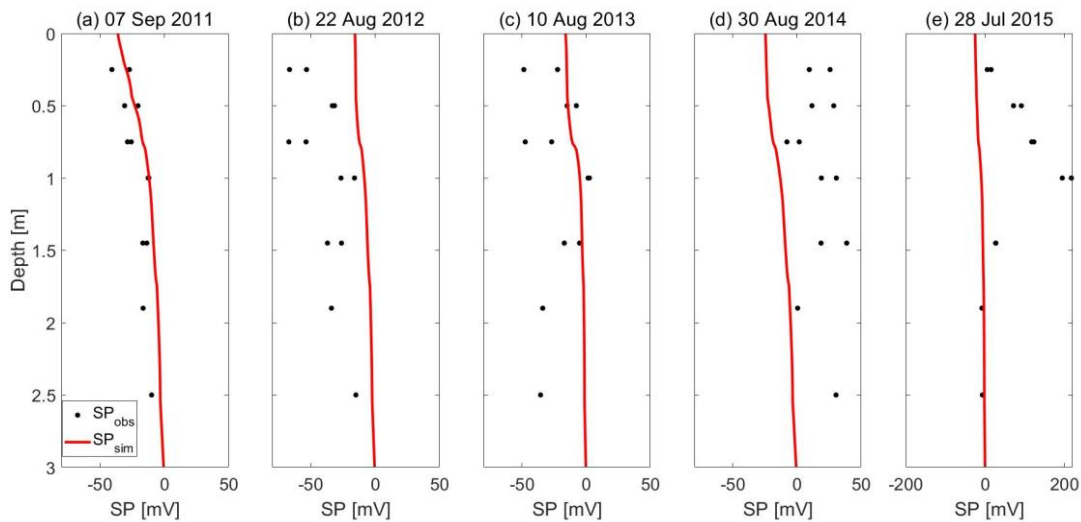

Fig. 8. Comparison between observed (black dots) and simulated (red lines) SP profiles on (a) 07 September 2011, (b) 22 August 2012, (c) 10 August 2013, (d) 30 August 2014 and (e) 28 July 2015 (e), respectively. These dates in during the late summer period of each year were chosen to correspond to days just during important rain events (25.32 mm, 33.77 mm, 18.66 mm, 22.48 mm and 37.74 mm, respectively).

*4.3. SP responses to infiltration events after years of monitoring*

Based on the simulation of solute transport presented in Fig. 7, the injected saline water should disappear from the vadose zone after April 2013. Figure 8 highlighted strong deviations from the simulated data starting a year from the electrode installation that we are unable to explain without invoking changes in the electrode potential (i.e., electrode drifts). The common approach when facing unexplained electrode offsets, is to refer to changes in the SP data with respect to a reference datum. This approach can only work satisfactorily if the "electrode drift" is slow and unrelated to the local water content and salinity. With the aim of highlighting the SP response to heavy rainfall events after more than three years of monitoring, we consider the SP data at 0.25 m and 1.9 m during three days in the summer of 2014 and the winter of 2015, respectively. We show the curves in Figs. 9 and 10 with different scales, as there are large differences in amplitudes between 0.25 m and 1.9 m. In Figs. 9 and 10, the SP data are presented as relative changes compared with the SP data recorded at the corresponding depth at 00:00:00 on August 30, 2014 and at 12:00:00 on January 7, 2015, respectively. This type of re-referencing of SP data is very common (e.g., Linde et al., 2007a). Furthermore, the reference electrode was set at 2.5 m to focus on temporal changes in the near surface. The periods were chosen to correspond to a long dry period followed by heavy rainfall (Fig. 9a and Fig. 10a) leading to strong changes in water content (Figs. 9b, d and Figs. 10b, d). In the summer season of 2014, one of the SP electrodes at 0.25 m show responses corresponding to the timing of changes in the water content. At 1.9 m, there is an SP response (Fig. 9e) with the same sign as in the winter (Fig. 10e), but the initiation of the SP signal occurs with a time delay with respect to the water content signal (Fig. 9d). Considering the winter period, the SP data at 0.25 m (Fig. 10c) do not show a clear unambiguous relationship to the rainfall event and the two SP electrodes respond rather differently. However, the SP data at 1.9 m (Fig. 10e) show as expected increasing negative amplitudes starting at a time that is only slightly earlier than the response in the water content at 2.0 m (Fig. 10d). These differences might suggest significant lateral variations in vertical flow.

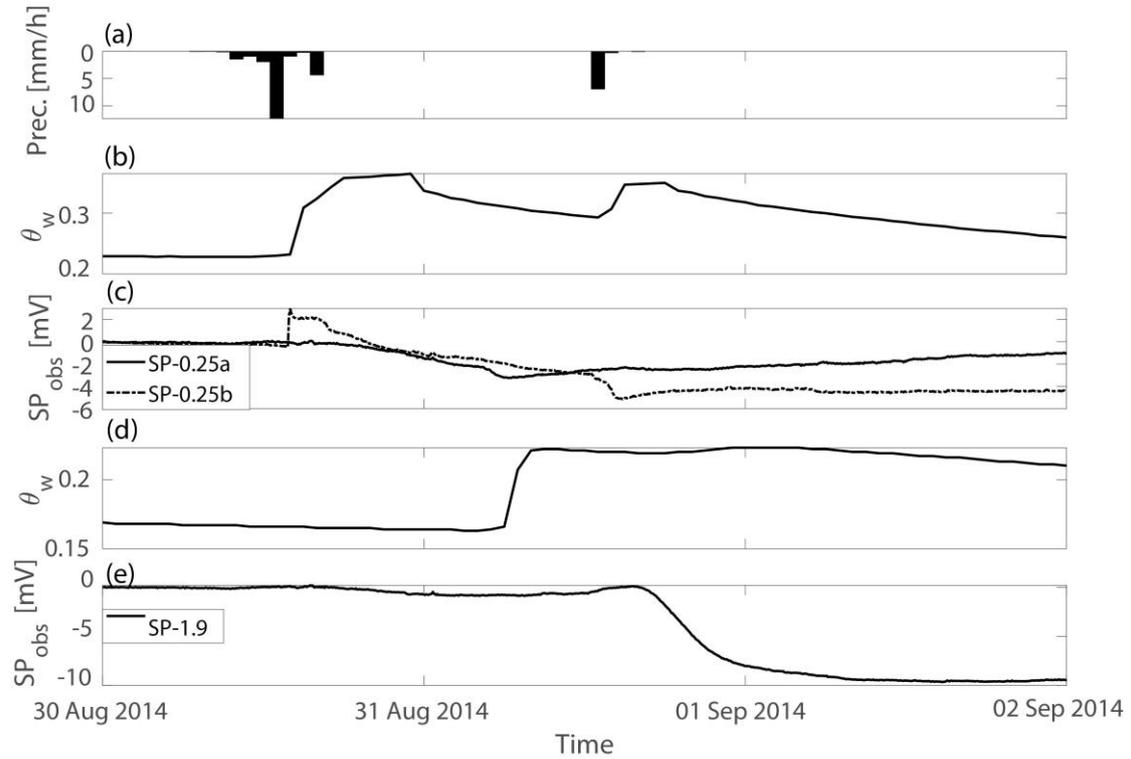

Fig. 9. SP responses observed during a period with significant rain events separated by dry periods in the summer of 2014. (a) Precipitation data; (b) measured water content data at 0.25 m; (c) observed SP data at 0.25 m; (d) measured water content data at 2 m and (e) observed SP data at 1.9 m. Here, the SP reference is located at 2.5 m depth and the amplitudes are set to zero at the beginning of the plotted period.

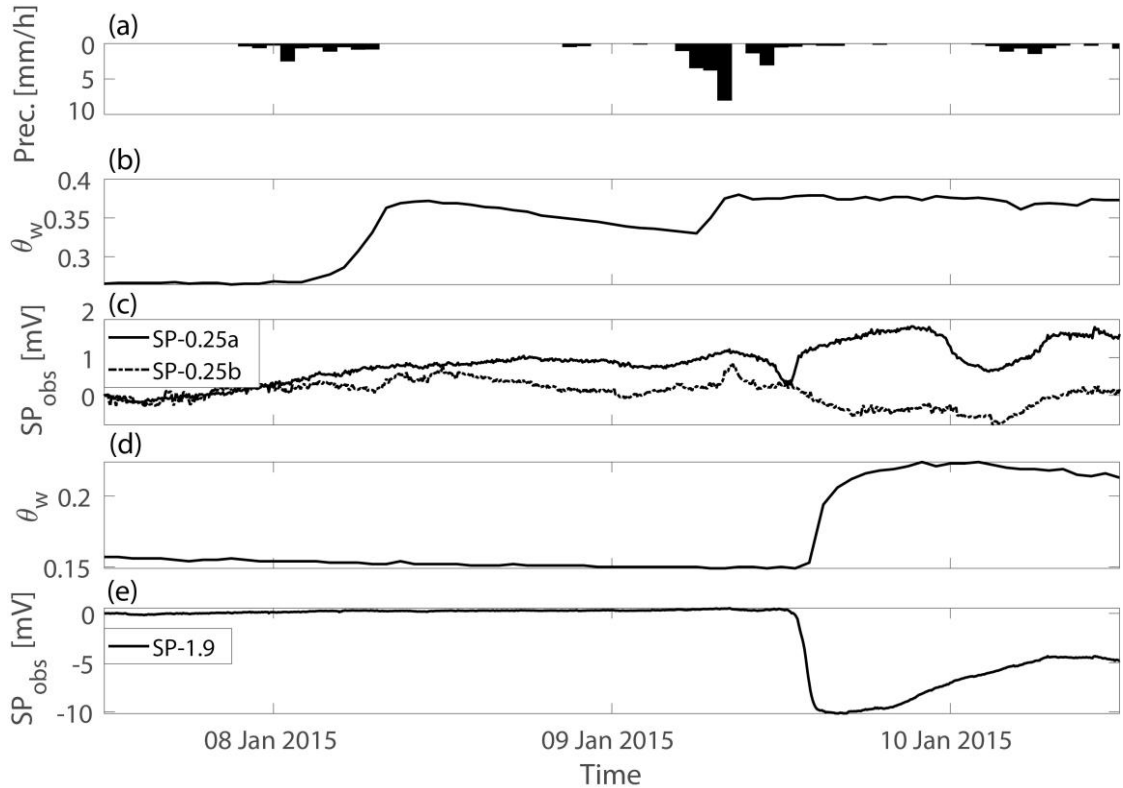

Fig. 10. SP responses observed during a period with significant rain events separated by dry periods in the winter of 2015. (a) Precipitation data; (b) measured water content data at 0.25 m; (c) observed SP data at 0.25 m; (d) measured water content data at 2 m and (e) observed SP data at 1.9 m. Here, the SP reference is located at 2.5 m depth and the amplitudes are set to zero at the beginning of the plotted period.

## 5. Discussion

### 5.1. SP-data with bentonite installation

The electrode surrounded by bentonite was installed several meters away from the saline injection area and is likely unaffected by the saline injection. The bentonite with its low permeability is likely to offer a stable environment in which water content and salinity variations are low. Indeed, the use of bentonite is often recommended for SP monitoring (Petiau, 2000). We consider here the SP data at 0.5 m, from the installation time until the saline tracer injection. In this time period, it is expected that the main signals are of electrokinetic origin. The overall shape of the SP electrode in bentonite (Fig. 11b) is similar to the numerical simulations, which ignores the presence of the bentonite (Fig. 11d). The two time-series show a stepped shape with declines occurring around July 22, August 7 and September 6 corresponding to the main rain events in this time-period. On top of this, the electrode in the bentonite shows a shift in magnitude and a linear drift with increasing negative values over time that is not present in the simulated data. One of the electrodes installed at 0.5 m (Fig. 11c) shows temporal variations corresponding to the same events, but the magnitudes are smaller than in the bentonite and the polarity of the response is not always the same (negative on July 22, negative on August 7 and positive on September 6).

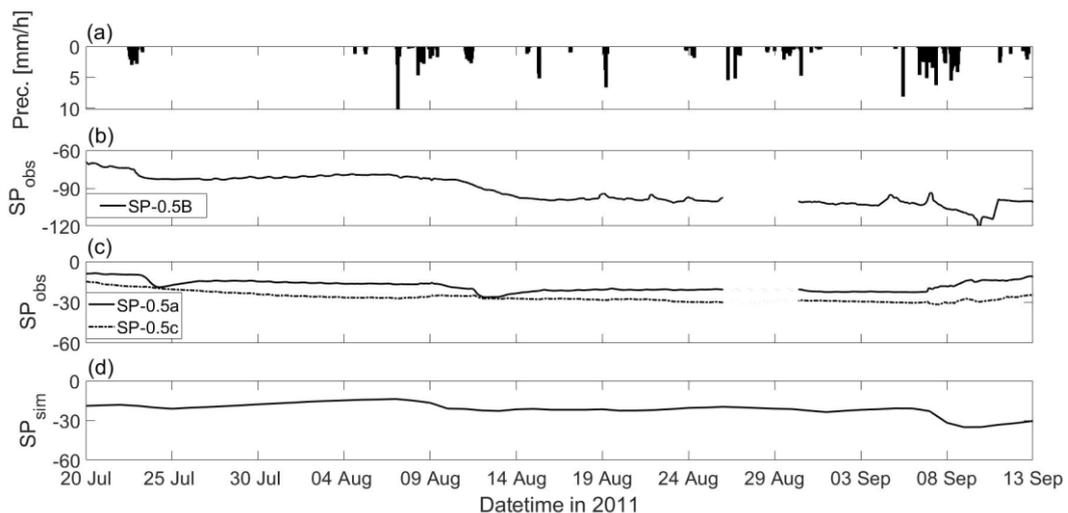

Fig. 11. (a) Precipitation data in the considered time period. Comparison between the observed SP signals in (b) bentonite with (c) observed SP signals without bentonite installation and (d) simulated SP ignoring the presence of the bentonite at 0.5 m. SP data were not recorded from 26 August 2011 to 30 August 2011.

A notable feature is clearly that the SP signals in the bentonite (Fig. 11b) are much smaller (-65 mV on average) than the other SP electrodes at the same depth (Fig. 11c) and the simulated SP data (Fig. 11d). This is likely related to the fact that bentonite is a microporous media with a much larger fraction of the pore space being affected by the electric double layers (EDL). One would thus expect anion exclusion effects (Jougnot et al., 2009; Leinov and Jackson, 2014) implying that Cl$^-$ is excluded from the pore space, leading to a different macroscopic Hittorf number $T_{Na}^{ben}$ in the bentonite compared to the surrounding. Considering the simple case for which the soil salinity is uniform, the difference of SP between the bentonite $SP_{ben}^{exc}$ and the surrounding soil $SP_{soil}^{ele}$ is:

$$\Delta\varphi_{ben}^{exc} = SP_{ben}^{exc} - SP_{soil}^{ele} \tag{32}$$

$$= -\frac{k_B T_K}{e_0}(2T_{Na}^* - 1)\ln\frac{C^{ele}}{C^{ben}} - \frac{k_B T_K}{e_0}(2T_{Na}^{ben} - 1)\ln\frac{C^{ben}}{C^{soil}}$$

$$+ \frac{k_B T_K}{e_0}(2T_{Na}^* - 1)\ln\frac{C^{ele}}{C^{soil}}$$

$$= \frac{k_B T_K}{e_0}(1 - 2T_{Na}^*)\ln\frac{C^{soil}}{C^{ben}} + \frac{k_B T_K}{e_0}(1 - 2T_{Na}^{ben})\ln\frac{C^{ben}}{C^{soil}}$$

where $C^{ele}$, $C^{ben}$ and $C^{soil}$ are the ion concentrations inside the electrode, in the bentonite and in the surrounding soil. Considering the limiting case $T_{Na}^{ben} = 1$, $\Delta\varphi_{ben}^{exc}$ could be -66.32 mV when $\frac{C^{ben}}{C^{soil}} = 10$ and $T_{Na}^* = 0.43$ at 20°C (293.15 K) in good agreement with the observations in Figs. 11b and 11c. This example suggests that the electrode installation method has a large effect on the measured SP signals (absolute amplitudes, drifts and sensitivity to perturbations) and that monitoring of ion concentrations or suitable proxies (e.g., electrical conductivity) in the vicinity of the electrodes are needed.

*5.2. Electrode effects*

One risk in the context of multi-year SP monitoring is that other ions than Cl⁻ that are naturally present in the soil enter into contact with the lead wire through diffusion (Petiau, 2000). This may lead to unpredictable perturbations of the electrode polarization that arise well before the desaturation time of Cl⁻ (Petiau, 2000). Judging from our data (Fig. 8), an even larger risk is related to dehydration of the electrolyte by capillarity occurring when the SP electrodes are installed under unsaturated conditions with the dehydration process varying over time and space as the capillary pressure varies. In contrast to the short-term laboratory findings by Petiau (2000) discussed in section 2.1.3, our long-term field-based results suggest that the lifespan of Petiau electrodes is greatly reduced in partially saturated media. We attribute this to capillary suction removing the electrolyte from the clay mud inside the electrodes. Petiau (2000) reports that the measured electrode potential increases when the electrodes dry out as a consequence of the limited contact between the remaining electrolyte and the lead wire inside the electrodes. This is in strong agreement with the observed shift towards positive values over time (Fig. 8). Our field observations suggest that Petiau electrodes installed in the vadose zone have considerably shorter life-times than previously suggested by Petiau (2000).

It could be argued that the impact of time-varying saturation and concentration state on the electrodes could be included in the modeling and that their impact on the measured SP data could be predicted. However, observed differences between electrodes located close to each other at the same depth are essentially unpredictable when considering 1-D flow and transport simulations. To characterize this aspect of electrode drift related to local variations in the electrode installation, geological heterogeneity and perhaps the electrodes themselves, we plot the median of the absolute deviation of the electrode pairs installed at common depths (Fig. 12a). The electrodes are stable with low absolute deviations before May 2012, after which the discrepancies increase and reach a peak of almost 40 mV around May 2013. It is also seen that the median of the absolute deviations is overall lower in the summers than in the winters. The largest absolute deviations are found for the shallow electrodes. This behavior is also seen in Fig. 5, in which the deviations among the electrode pairs increase notably after May 2012. The median value of the absolute deviation between simulated water content and measured water content inverted by measured dielectric constant data using Eq. (27) shows similar seasonal trends (Fig. 12b). Figure 12b offers, thus, some possible insights regarding the possible causes of the larger absolute deviations in the observed SP data (Fig. 12a) during the winter period. Indeed, partial freezing of the soil and snow accumulation on the surface lead to heterogeneous melt water infiltration (making the underlying 1-D assumption increasingly questionable) and various measurement issues (Vásquez et al., 2015).

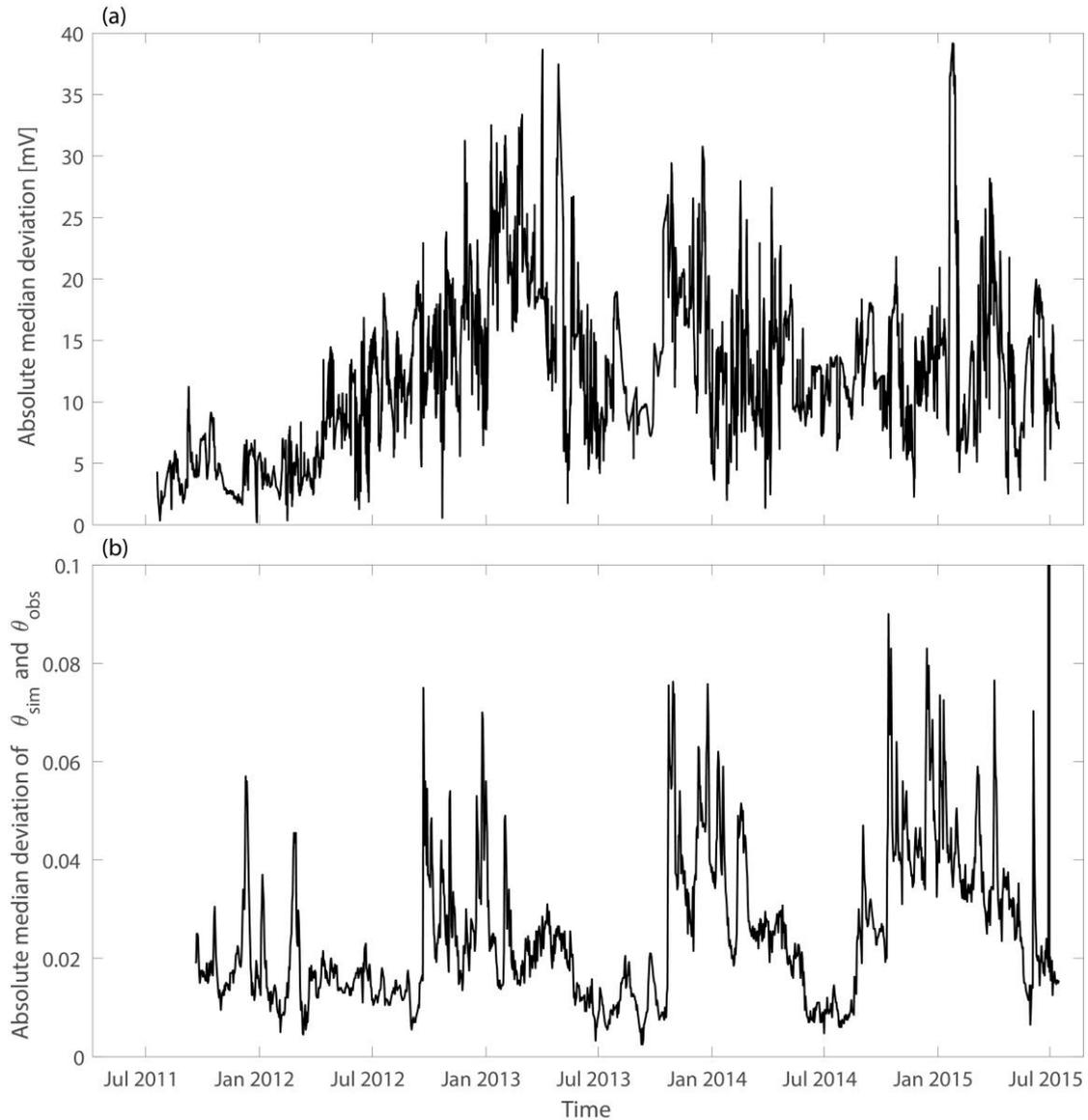

Fig. 12. (a) Temporal variation of the absolute median deviation of observed SP data at depths (0.25 m, 0.5 m, 0.75 m, 1 m, 1.45 m) with two measurement electrodes. The corresponding absolute median deviations of each electrode pair for the whole time period are 15.08 mV, 12.92 mV, 9.03 mV, 9.04 mV, 8.96 mV, respectively. (b) Temporal variation of the absolute median deviation of observed water content data with simulated water content data at depths (0.25 m, 1 m, 1.5 m, 2 m, 2.25 m, 2.5 m, 2.75 m, 3 m). The corresponding absolute median deviations at each depth over time are 0.03, 0.02, 0.03, 0.01, 0.02, 0.02, 0.02, 0.02, respectively.

## 6. Concluding remarks

The analysis of four years of SP monitoring data at the HOBE hydrological observatory in Denmark has highlighted that modeling of streaming potentials in the vadose zone is still a largely unresolved challenge. The main complication is the strong variability of the effective excess charge ($\widehat{Q^{\text{eff}}}(\theta_w, C_w)$) and the difficulties in predicting this function accurately. Traditionally, it is expected that streaming potential magnitudes increase when water fluxes increase (e.g., Doussan et al., 2002). However, this does not need to be the case and increased flow might actually lead to decreases in the observed SP amplitudes (see Figs. 4f and 4g). As the water content increases, $\widehat{Q^{\text{eff}}}(\theta_w, C_w)$ decreases particularly at the onset of fluid flow in macro pores (these pores carry a lot of flow without a significant drag of excess charge). So, this decrease in $\widehat{Q^{\text{eff}}}(\theta_w, C_w)$ might offset the increase in current density offered by the increasing water flux **u** (Eq. 5), while the increasing water content increases electrical conductivity (Eq. 3), thereby, decreasing the observed self-potential (Eq. 2). Calibration of such non-linear functions as $\widehat{Q^{\text{eff}}}(\theta_w, C_w)$ to achieve high predictive capability is very challenging. Dual-domain formulations of porosity need to be considered to account for the influence of macro pores on flow, transport and electrical properties (see also discussion Romero-Ruiz et al. 2019). This work has also highlighted the need to accurately model the time-evolving water salinity, which implies that water-soil interactions need to be accounted for, and that water salinity needs to be monitored. Furthermore, it is probably also needed to account for the fact that flux-averaged soil water concentrations are different from in-situ concentrations (e.g., Evaristo et al., 2015). This implies that predictive SP modeling in the vadose zone does not only need to consider advanced SP theory, but also state-of-the-art flow and transport modeling accounting for chemical soil interactions in a dual- or multi-continuum framework. Our work clearly highlights that SP signals in the vadose zone cannot be interpreted as a sensor of water movement by simply differencing SP differences between neighboring depth levels. The reality is much more complicated than this even in absence of electrode effects.

Our study confirms suggestions by Perrier et al. (1997) and Doussan et al. (2002) that the design and installation of SP-electrodes for long-term monitoring in the vadose zone is a largely open topic. Important electrode-related challenges in predictive SP modeling concern (*i*) the membrane polarization between the interior of the electrode and its immediate surroundings and (*ii*) dehydration of the electrodes by capillary forces impacting electrode potentials. Except for one electrode, we did not follow the common approach of installing the electrodes in a bentonite solution (Perrier et al., 1997; Petiau, 2000; Doussan et al., 2002). Clearly, a bentonite solution improves the contact resistance with the ground and makes the data less sensitive to small-scale heterogeneity (Petiau, 2000). However, it also leads to difficult-to-predict electro-diffusive effects as the bentonite equilibrates with the surrounding media and the electrode solution leading in our case to large electrode offsets with respect to other electrodes and enhanced drifts (Fig. 11b). For instance, the soil mud in which Doussan et al. (2002) inserted their electrodes displayed a threefold decrease in the pore water electrical conductivity during the monitoring period and, thus, a similar decrease in salinity. This implies significant electro-diffusive effects of 3-D nature in the soil and does not ensure constant water salinity at the contact with the electrode. For future studies, we strongly recommend that pore water conductivity and chemistry is monitored, which would help to elucidate some of the behaviors that we report.

**Acknowledgments**

We thank the Danish hydrological observatory HOBE for providing access to the site, technical help, and full access to the data. The data used in this paper are available here: http://dx.doi.org/10.17632/6r8898657w.1. To calculate the SP signals, we used a version of Maflot (maflot.com) that was kindly provided by I. Lunati and R. Künze and modified by M. Rosas-Carbajal. We are grateful for the financial support offered by the China Scholarship Council and the helpful reviews by Emily Voytek and André Revil.